\documentstyle[prl,multicol,aps,epsfig]{revtex}

\begin{document}
\title{Anisotropic charge displacement supporting isolated photorefractive optical needles}
\author{Eugenio DelRe}
\address{Advanced Research, Pirelli Cavi e Sistemi, Viale Sarca 222, 20126 Milan, Italy}
\author{Alessandro Ciattoni}
\address{Dipartimento di Fisica, Universit$\grave{a}$ dell'Aquila, and INFM, Unit$\grave{a}$ di Roma 1, Rome, Italy}
\author{Aharon J. Agranat}
\address{Department of Applied Physics, Hebrew University of Jerusalem, Jerusalem 91904, Israel}
\date{\today}
\maketitle

\begin{abstract}
The strong asymmetry in charge distribution supporting a single
non-interacting spatial needle soliton in a paraelectric
photorefractive is directly observed by means of
electroholographic readout.  Whereas in trapping conditions a
quasi-circular wave is supported, the underlying double-dipolar
structure can be made to support two distinct propagation modes.

\end{abstract}

\pacs{??????????}

\begin{multicols}{2}
\narrowtext  Far from being a peculiarity of low dimensional
systems, solitary waves and solitons have been widely documented
in bulk three-dimensional environments
\cite{general-spatial-soliton}. In biased photorefractives,
nonlinear visible optical waves have been shown to undergo
self-trapping both as extended one-dimensional waves, in the form
of slab-solitons \cite{first-paper}, and as confined
two-dimensional spatial beams, needle-solitons
\cite{needle-solitons}. These are self-funneled micron-sized beams
of light that propagate through the bulk dielectric without
suffering diffraction or distortion. Needles, in their richer
higher-dimensional environment, have led to a substantial advance
in our phenomenological investigation of nonlinear dynamics,
expanding the scope of possible soliton-based applications
\cite{applications} \cite{electro-optic}.

Whereas both slabs and needles emerge in the same physical system,
a biased photorefractive sample, their underlying nonlinear
nature is rather different \cite{theory1}. For slabs, the entire
physical system, and thus, consequently, the optical
nonlinearity, depends only on the transverse beam direction along
which the external field is applied (say the x direction),
whereas the system is fully invariant for spatial translations in
the second orthogonal transverse direction y. This reduces slab
soliton description to that associated with a saturated Kerr-like
nonlinearity \cite{screening-theory}.  For needles, on the
contrary, the higher dimensionality of the optical beam, whose
quasi circular symmetry suggests an isotropic self-action
\cite{needle-solitons}, is inherently at odds with the screening
nonlinearity, whose one basic driving mechanism is the x directed
external bias field.  A simplified description of needles,
tracing the steps that lead to a local Kerr-like understanding of
slabs, is simply not possible \cite{theory2}. Given the
complexity of the higher-dimensional interaction, the theoretical
interpretation of needles is largely based on numerical
integration.  What emerges is a picture in which nonlocal
nonlinear effects \cite{theory3}, as opposed to local
conventional paradigm Kerr-like phenomenology, play a central
role. An understanding of these requires an explicit distinction
between the underlying space-charge field distribution $E_{sc}$,
which mediates self-action, and the propagating light field
$E_{opt}$. The space-charge distribution simply does not have a
{\it local} relationship to the optical field \cite{theory2}. The
numerical solution of the full boundary-value problem indicates
that the highly anisotropic screening configuration allows the
formation of needles only through an equally anisotropic local
space-charge, characterized by the appearance of two distinct
lateral field lobes in the x direction, absent in the second
transverse direction y \cite{theory1} \cite{theory3}. This double
dipolar field distribution induces, as a consequence, a
complicated needle supporting index pattern that has little to do
with a mere self-written graded-index waveguide (excluding the
possibility of a simple linear interpretation \cite{snyder}). For
system parameters far from the soliton supporting configuration,
this anisotropy leads to an observable asymmetric beam
distortion, but the question naturally arises as to how these
lobes manifest their existence when the optical beam is actually
a needle-like solitary wave.

Repulsion of mutually incoherent needles provides indirect
evidence of the lobe-like charge distribution \cite{anomalous}.
However, no direct experimental evidence of charge anisotropy has
yet been reported.  The main reason lies in the fact that
photorefractive solitons are generally observed in ferroelectric
samples.  In these crystals there is no direct way of isolating
the contribution of charge displacement from the final guiding
structure.  Read-out with non-photorefractively active light can
lead to no substantial increase in knowledge on the underlying
charge pattern, short of performing precise bulk interferograms
or far field soliton transforms \cite{theory3}.  Direct
investigation of the space-charge residue with a probe is
furthermore hampered by the fact that the lobes are actually {\it
antiguiding} \cite{theory1} \cite{theory3}.

In this Letter we give direct evidence of this nonlocal field
structure.  This is made possible by the quadratic electro-optic
response of paraelectrics, that allows the electro-holographic
separation of optical phenomenology from the underlying
space-charge field \cite{electro-optic}.

Experiments are carried out in a sample of photorefractive
3.7$^x$x4.7$^y$x2.4$^z$ mm KLTN
(potassium-lithium-tantalate-niobate) \cite{ronnie-crystal} ,
biased along the x axis (of size L=3.7mm), and kept at a constant
temperature T=20$^{\circ}$C. The x-polarized cw TEM$_{00}$
$\lambda$=532nm beam from a diode-pumped doubled NdYag laser is
focused on the input facet of the sample and launched along the z
axis. As it propagates in the sample, it diffracts, passing from
an initial intensity I=$|E_{opt}|^2$ full-width-half-maximum
(FWHM) in the x and y directions $\Delta x \cong \Delta y \cong 10
\mu m$ to a broadened intensity distribution of $\Delta x \cong
\Delta y \cong 20 \mu m$ (see Fig1.(a),(b)). The application of
the external constant bias V on the x electrodes makes
photoexcited free charges drift, leading to an inhomogeneous
field screening.  The electro-optic response of the paraelectric
sample is $\Delta n = -(1/2)n^3 g_{11} \epsilon_0^2
(\epsilon_r-1)^2(V/L)^2 (E/(V/L))^2 \equiv -\Delta n_0
\mathcal{E}$ $^2$, where $n \cong$ 2.4 is the zero-field index of
refraction, $g_{11} \equiv g_{xxxx} \cong$ 0.12m$^4C^{-2}$ is the
dominant component of the quadratic electro-optic tensor
g$_{ijkl}$ (and thus tensorial effects are neglected),
$\epsilon_0$ is the vacuum dielectric constant, $\epsilon_r
\cong$ 9$\cdot$ 10$^3$ (at T$\cong$20$\circ$C) is the relative
sample low frequency dielectric constant, $E$ is the x component
of the electric field resulting from screening, $\Delta n_0 \cong
2.8\cdot10^{-4}$, and $\mathcal{E}$ $\equiv E/(V/L)$. The
spatially modulated index distribution allows needle formation
(see Fig.1(c)).  The needle, that shows a slight anisotropy in
the output intensity distribution, is trapped and stable in time
for an external bias voltage of V=0.85kV and a ratio of peak
intensity I$_p$ to the dark artificial illumination I$_b$
(obtained by illuminating the sample with a copropagating y
polarized plane wave of equal wavelength) of u$_0^2 \equiv
I_p/I_b \cong$2.6.  Annulling the externally applied voltage V,
i.e., setting V=0, gives an index modulation $\Delta n_{V=0} =
-\Delta n_0 \mathcal{E}$ $_{sc}^2$, {\it only} due to the charge
displacement, where evidently $\mathcal{E}$ $_{sc} \equiv$
$\mathcal{E}$-1.  The resulting index pattern has a guiding
structure for regions in which $\mathcal{E}$ $_{sc}$ passes
through a {\it minimum}. Given that the lobes represent an excess
of screening in the x direction \cite{theory1} \cite{theory3},
there are two points, i.e., x$_1$ and x$_2$, along the x axis,
located to the left and right of the needle peak, in which
$\Delta n_{V=0}$ forms a guiding "hump". Along the y axis, this
hump will follow the shape of the lobe.

In order to investigate $\Delta n_{V=0}$ {\it without} modifying
the space-charge distribution, we launch into the sample the same
beam leading to the needle, but attenuated so as to have a much
lower intensity.  This guarantees that the characteristic time
scale of charge displacement induced by the probe, $\tau_d$, is
much longer than any characteristic observation time.  For
typical $\mu$W intensity beams, $\tau_d \sim$1 min.

Results, shown in Fig.1(d), clearly indicate the anisotropic lobe
structure in the form of a split diffraction pattern in the x
direction. The slight asymmetry in the diffraction pattern is a
consequence of needle self-bending, that inevitably distorts the
diffractive read-out phase.

\begin{figure}
\begin{center}
\resizebox{8.5cm}{!}{\mbox{\includegraphics*[0cm,3cm][20cm,27cm]{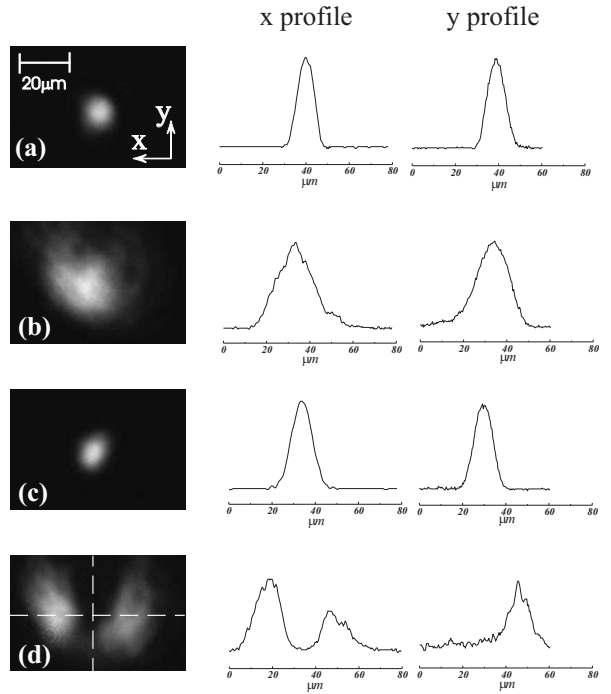}}}
\caption{Electroholography of a single photorefractive needle.
(a) Image and profiles of input transverse intensity distribution;
(b) Linear diffraction with nonlinear charge separation turned off
(V=0); (c) Self-trapping distribution for V=0.85kV; (d) Read-out
for V=0.} \label{zero-voltage-steady-state}
\end{center}
\end{figure}

A similar phenomenology has been observed for transient
quasi-steady-state needles, where I$_b$=0, blocking beam
evolution in the trapped regime, i.e., before the needle has
decayed.

The two light lobes are a signature of the lobes predicted by
numerical integration of the full Kukhtarev model and constitute
direct proof that the nonlinearity supporting needle trapping in
biased photorefractives is not the saturated Kerr-like $\Delta n
\propto 1/(1+I/I_b)^2$ that allows slab formation.  More
precisely, whereas the lobes are {\it not} present in the slab
case (and are not merely "negligible"), they play a fundamental
role in needle trapping \cite{anisotropy}. Although needles have
been documented in various conditions, it is legitimate to ask
whether the nonlocal space-charge field structure, and thus index
modulation, can actually support circular-symmetric solitary
waves. The mathematical answer is no \cite{theory4}.  However,
the anisotropic space-charge structure {\it can} support waves
that are to all practical purposes circular-symmetric.  For the
conditions investigated experimentally, we find the space-charge
distribution by solving the simplified associated electro-static
problem, i.e., $\nabla \cdot [(I+I_b) {\bf E}+(K_bT/q)\nabla
I]$=0, where ${\bf E}$ \cite{theory2} is the internal electric
field vector, assuming a given Gaussian intensity distribution.
The resulting index pattern is shown in Fig.(2a). Propagating the
very same field distribution $E_{opt}$ (whose intensity is I,
shown in Fig.(2b)) into this pattern, gives results shown in
Fig.(2c)-(2d).  The intensity pattern does not suffer discernible
distortion.  This means that the exact nonlinear behaviour is
well described by this approximate linear approach, and thus we
can conclude that quasi-circular needles can be supported by the
anisotropic pattern.

\begin{figure}
\begin{center}
\resizebox{8.5cm}{!}{\mbox{\includegraphics*[2cm,1cm][28cm,19cm]{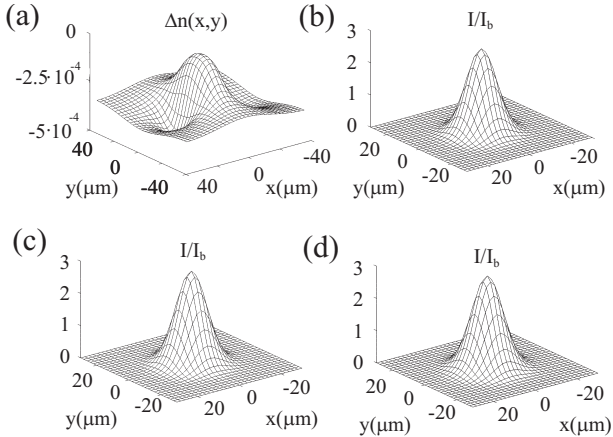}}}
\caption{Self-consistency of a needle solitons trapped in the
anisotropic nonlinear index pattern. (a) Anisotropic index
pattern; (b) input intensity distribution; (c)-(d) Intensity after
2mm and 4.5mm propagation, respectively, for the experimental
situation described above.} \label{anisotropy}
\end{center}
\end{figure}

One basic consequence of these findings is that the anisotropy
underlying a photorefractive needle leads not to one, but to {\it
three} spatially separated index structures, that can be made to
alternatively guide light depending on the applied external
voltage in the read-out phase.  This would not have been possible
had the nonlinear response been local, as in the one-dimensional
case \cite{electro-optic}.  The electroholographic read-out would
have implied a transition from a localized single mode structure
(the needle) to a delocalized "doughnut-like" guiding pattern. To
demonstrate this, we investigate the guiding capabilities at V=0.
We were able to show the two guided modes launching, in sequence,
the probe beam into one of the two lateral guiding humps of the
$\Delta n_{V=0}$ pattern, i.e., in $x_1$ and $x_2$. Results are
shown in Fig.3.  We did not observe any directional coupling
between the modes, this clearly being a consequence both of the
distance between the humps, the probe wavelength, propagation
length, and the presence of the antiguiding central pattern.

\begin{figure}
\begin{center}
\resizebox{8.5cm}{!}{\mbox{\includegraphics*[1cm,3cm][29cm,18cm]{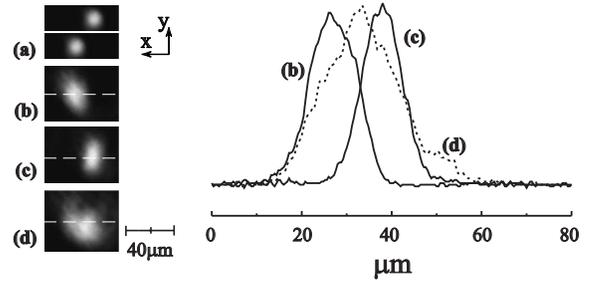}}}
\caption{Double hump guiding structure.  (a) Two different input
beams, shifted by approximately $\pm$10$\mu$m; (b-c) Guided beam
in the two humps; (d) Linear diffraction for unshifted beam.}
\label{lobe-guiding}
\end{center}
\end{figure}

The work of E.D. was partially carried out during previous
activity at Fondazione Ugo Bordoni. The work of A.C. was funded by
the Istituto Nazionale Fisica della Materia through the PAIS2000
SESBOM project. Research carried out by A.J.A. is supported by a
grant from the Ministry of Science of the State of Israel.

\end{multicols}


\begin{thebibliography} {aa}
\bibitem{anisotropy} There is no fundamental
difference between anisotropy in paraelectrics and ferroelectrics.
\end{thebibliography}
\end{document}